\documentclass{article}

\usepackage{microtype}
\usepackage{graphicx}
\usepackage{subfigure}
\usepackage{booktabs} % for professional tables

\usepackage{hyperref}

\usepackage[accepted]{icml2023}

\usepackage{amsmath}
\usepackage{amssymb}
\usepackage{mathtools}
\usepackage{amsthm}
\usepackage{stfloats}
\usepackage{bm}
\usepackage{algorithm}
\usepackage{algorithmic}
\usepackage{bbding}
\usepackage{pifont}
\usepackage{dsfont}
\usepackage{caption}

\usepackage[capitalize,noabbrev]{cleveref}

\theoremstyle{plain}

\theoremstyle{definition}

\theoremstyle{remark}

\newcommand{\ptheta}{p_\theta}

\usepackage[textsize=tiny]{todonotes}

\icmltitlerunning{}

\begin{document}

\twocolumn[
\icmltitle{ DiffSDS: A language diffusion model for protein backbone inpainting under geometric conditions and constraints}

% Inserting direction space into language diffusion models for conditional protein backbone generation under constraints



\icmlsetsymbol{equal}{*}

\vspace{-3mm}
\begin{icmlauthorlist}
\icmlauthor{Zhangyang Gao}{equal,westlake,zhejiang}
\icmlauthor{Cheng Tan}{equal,westlake,zhejiang}
\icmlauthor{Stan Z. Li}{westlake}

\end{icmlauthorlist}

\icmlaffiliation{westlake}{AI Research and Innovation Lab, Westlake University}
\icmlaffiliation{zhejiang}{Zhejiang University}

\icmlcorrespondingauthor{Stan Z. Li}{Stan.ZQ.Li@westlake.edu.cn}

\icmlkeywords{Machine Learning, ICML}

\vskip 0.1in
]



\printAffiliationsAndNotice{\icmlEqualContribution} % otherwise use the standard text.

\begin{abstract}
    Have you ever been troubled by the complexity and computational cost of SE(3) protein structure modeling and been amazed by the simplicity and power of language modeling? Recent work has shown promise in simplifying protein structures as sequences of protein angles; therefore, language models could be used for unconstrained protein backbone generation. Unfortunately, such simplification is unsuitable for the constrained protein inpainting problem, where the model needs to recover masked structures conditioned on unmasked ones, as it dramatically increases the computing cost of geometric constraints. To overcome this dilemma, we suggest inserting a hidden \textbf{a}tomic \textbf{d}irection \textbf{s}pace (\textbf{ADS})  upon the language model, converting invariant backbone angles into equivalent direction vectors and preserving the simplicity, called Seq2Direct encoder ($\text{Enc}_{s2d}$). Geometric constraints could be efficiently imposed on the newly introduced direction space. A Direct2Seq decoder ($\text{Dec}_{d2s}$) with mathematical guarantees is also introduced to develop a \textbf{SDS} ($\text{Enc}_{s2d}$+$\text{Dec}_{d2s}$) model. We apply the SDS model as the denoising neural network during the conditional diffusion process, resulting in a constrained generative model--\textbf{DiffSDS}. Extensive experiments show that the plug-and-play ADS could transform the language model into a strong structural model without loss of simplicity. More importantly, the proposed DiffSDS outperforms previous strong baselines by a large margin on the task of protein inpainting.

\end{abstract}

\vspace{-3mm}
\section{Introduction}
We aim to improve and simplify the modeling of constrained protein backbone inpainting, i.e., recovering masked protein backbones, which has wide applications in \textit{de-novo} protein design \cite{wang2022scaffolding, lee2022proteinsgm, ferruz2022sequence}. Since the milestone breakthrough of AlphaFold \cite{jumper2021highly}, protein structure models are becoming increasingly sophisticated, including the introduction of equivalent learning biases, geometric interactions, and fine-grained structure modules \cite{ganea2021independent, jin2021iterative, luoantigen, lee2022proteinsgm, wang2022scaffolding, trippe2022diffusion, anand2022protein}. These improvements achieve significant success in structure modeling; however, the additional computational overhead and complexity have also troubled researchers. Despite the complexity of protein structure modeling, language transformers seem to unify everything for more difficult NLP tasks while enjoying good efficiency and simplicity. Can we convert constrained protein structure design as a sequence modeling task and thus apply language models for simplification?

Existing protein structural generative models explicitly consider the equivalence caused by rotation and translation \cite{wang2022scaffolding, trippe2022diffusion, luoantigen, anand2022protein}. These considerations enable them to correctly consider atom interactions in the 3D space while requiring special model designs that increase the complexity and computational cost. Restricted by the equivalence, traditional powerful models, such as visual CNNs or language converters, are prevented from being directly applied to structure modeling.  Beyond this limitation, recent FoldingDiff \cite{wu2022protein} suggests converting protein structures into sequences of angles. Thus, language models could be used for unconditional protein backbone generation. 

%画个图
\vspace{-3mm}
\begin{figure}[H]
    \centering
    \includegraphics[width=2.5in]{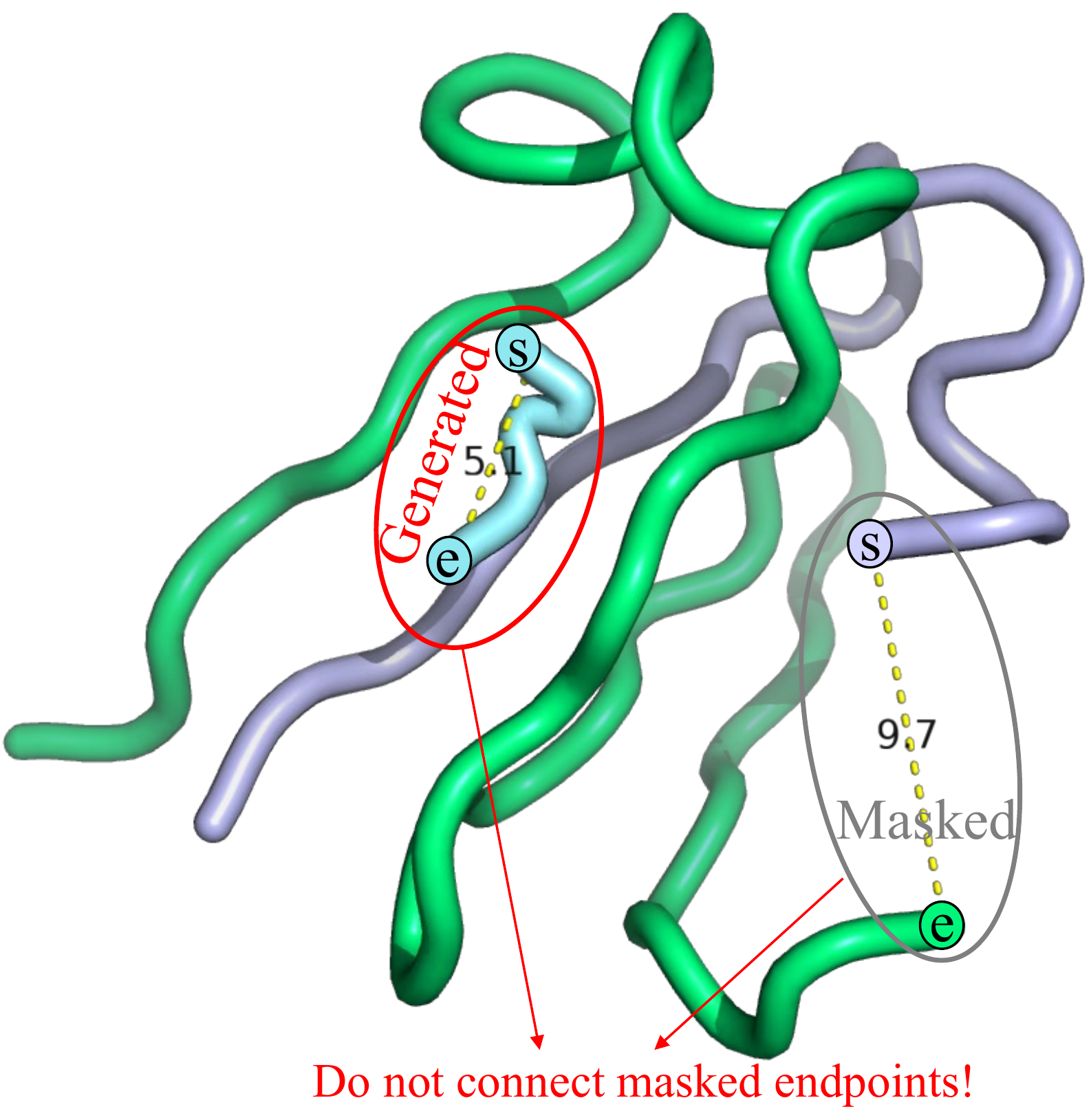}
    \vspace{-4mm}
    \caption{\small Example of violating geometric constraints, where endpoints $s$ and $e$ are not connected. The inputs are fixed unmasked atoms (colored with green), where atoms between endpoints $s$ and $e$ are masked and should be recovered by the algorithm. }
    \label{fig:break}
    \vspace{-3mm}
\end{figure}

Unfortunately, the pure language model is unsuitable for constrained structure design tasks. In protein backbone inpainting, the designed structure should fit multiple geometric constraints, including linking the masked structure's endpoints and not overlapping with unmasked structures to meet the repulsion \cite{spassov2007dominant, muller2010charge, drake2020physical}. If the above constraints are not guaranteed, the generated structure will be meaningless, as shown in Fig.\ref{fig:break}. An open research question is how to efficiently impose geometric constraints on the language model while keeping simplicity.

We suggest inserting a hidden \textbf{a}tomic \textbf{d}irection \textbf{s}pace (\textbf{ADS})  into the language model, allowing to impose structural constraints on the novel direction space efficiently. ADS is a plug-and-play cross-modal conversion technique connecting the sequence and direction space. By adding ADS upon the last transformer layer, we obtain the Seq2Direct encoder ($\text{Enc}_\text{s2d}$) that converts sequential features into direction vectors. We also introduce Direct2Seq decoder ($\text{Dec}_\text{d2s}$) according to strict mathematical transformations. A language model (SDS) equipped with hidden direction space could be constructed by stacking $\text{Enc}_\text{s2d}$ and $\text{Dec}_\text{d2s}$. SDS takes angular sequences as inputs and outputs, with a latent direction space that supports efficient geometric calculations and is mathematically consistent with the sequence space. In this design,  multimodal constraints, e.g., sequential and 3D constraints, can be simultaneously considered on the corresponding feature space. Finally, we apply the SDS model as the denoising neural network during the conditional diffusion process, resulting in a constrained generative model--DiffSDS.

We evaluated DiffSDS on CATH4.3 and compared it with recent strong baselines, including RFDesign \cite{wang2022scaffolding} and modified FoldingDiff \cite{wu2022protein}. We also propose three metrics to evaluate the results of protein backbone inpainting, including protein likeness, connectivity, and non-overlapping. Experiments show that our methods significantly outperform baselines in all metrics. In addition, the designability of structures generated by DiffSDS is also better than baselines. As to simplicity, the proposed DiffSDS utilizes the language transformer to model protein structures, avoiding the equivalence consideration when constructing neural modules. To provide a deeper understanding, we have also performed ablation studies to reveal the role of conditions and constraints.

\vspace{-3mm}
\subsection{Related work}

\paragraph{Problem Definition} Protein backbone inpainting aims to recover the continuous masked substructure of the protein backbone, given the unmasked atoms as conditions. The generated structure is required to connect different protein fragments with fixed spatial positions. Formally, we write the protein backbone as $\mathcal{B} = \{ \boldsymbol{p}^C_1, \boldsymbol{p}^A_1, \boldsymbol{p}^N_1, \boldsymbol{p}^C_{2}, \boldsymbol{p}^A_2, \boldsymbol{p}^N_2, \cdots, \boldsymbol{p}^C_n, \boldsymbol{p}^A_n, \boldsymbol{p}^N_n \}$, where $\{\boldsymbol{p}^C_i,  \boldsymbol{p}^A_i, \boldsymbol{p}^N_i\}$ indicates the set of backbone atoms ($C$, $C_{\alpha}$ and $N$) of the $i$-th residue and $\boldsymbol{p}^A_i$ is the 3D position of the $i$-th $C_{\alpha}$. Denote the masked sub-structure as $\mathcal{M} = \{ \boldsymbol{p}^C_i,  \boldsymbol{p}^A_i, \boldsymbol{p}^N_i \}_{i=s}^{e}$, the unmasked structures as $ \mathcal{K} = \{ \boldsymbol{p}^C_i,  \boldsymbol{p}^A_i, \boldsymbol{p}^N_i\}_{i=1}^{s-1}  \cup \{ \boldsymbol{p}^C_i,  \boldsymbol{p}^A_i, \boldsymbol{p}^N_i\}_{i=e+1}^{n}$, where $0<l<r<n$. We generate $\hat{\mathcal{M}}$ connecting endpoints $\boldsymbol{p}^{N}_{s-1}$ and $\boldsymbol{p}^C_{r+1}$ via a learnable function $f_{\theta}$, given $mathcal{U}$ with a fixed conformation as input:

\vspace{-4mm}
\begin{equation}
    \hat{\mathcal{M}} = f_{\theta}(\boldsymbol{p}| \mathcal{U}, x), x\sim \mathcal{N}(0,\mathbf{I})
\end{equation}
\vspace{-4mm}

Note that $\theta$ indicates learnable parameters. The designed $\hat{\mathcal{M}}$ should be non-trivial to satisfy the following constraints:

\vspace{-4mm}
\begin{enumerate}
    \item Protein likeness: The designed structures are likely to constitute natural proteins.
    \vspace{-2mm}
    \item Connectivity: $\hat{\mathcal{M}}$ should effectively connect $\boldsymbol{p}^{N}_{s-1}$ and $\boldsymbol{p}^C_{e+1}$ without breakage.
    \vspace{-2mm}
    \item Non-overlapping: The designed structure $\hat{\mathcal{M}}$ should not overlap with existing structure $\mathcal{U}$.
    \vspace{-2mm}
\end{enumerate}

\vspace{-2mm}
\paragraph{3D Molecule Generation} Generating 3D molecules to explore the local minima of the energy function (Conformation Generation) \cite{gebauer2019symmetry, 10.5555/3524938.3525768, simm2020reinforcement, shi2021learning, xu2021end, luo2021predicting, xu2020learning, ganea2021geomol, xu2022geodiff, hoogeboom2022equivariant, jing2022torsional, zhu2022direct} or discover potential drug molecules binding to targeted proteins (3D Drug Design)  \cite{imrie2020deep, nesterov20203dmolnet, luo20223d, ragoza2022chemsci, wu2022diffusion, huang2022mdm, peng2022pocket2mol, huang20223dlinker, wang2022generative, liu2022generating} have attracted extensive attention in recent years. Compared to conformation generation that aims to predict the set of favourable conformers from the molecular graph, 3D Drug Design is more challenging in two aspects: (1) both conformation and molecule graph need to be generated, and (2) the generated molecules should satisfy multiple constraints, such as physical prior and protein-ligand binding affinity. We summarized representive works of 3D drug design in Table.\ref{tab: mol_generate_models} in the appendix, where all the methods focus on small molecule design.

\vspace{-2mm}
\paragraph{Protein Design} In addition to small molecules, biomolecules such as proteins have also attracted considerable attention by researchers \citep{ding2022protein, ovchinnikov2021structure, gao2020deep, strokach2022deep}. We divide the mainstream protein design methods into three categories: protein sequence design \citep{li2014direct, wu2021protein, pearce2021deep, ingraham2019generative, jing2020learning, tan2022generative, gao2022alphadesign, hsu2022learning, dauparas2022robust, gao2022pifold, o2018spin2, wang2018computational, qi2020densecpd, strokach2020fast, chen2019improve, zhang2020prodconn, anand2022protein}, unconditional protein structure generation \citep{anand2018generative, sabban2020ramanet, eguchi2022ig, wu2022protein}, and conditional protein design \citep{lee2022proteinsgm, wang2022scaffolding, trippe2022diffusion, lai2022end, fu2022antibody, tischer2020design, anand2022protein, luoantigen}. Protein sequence design aims to discover protein sequences folding into the desired structure, and unconditional protein structure generation focus on generating new protein structures from noisy inputs. We are interested in conditional protein design and consider multiple constraints on the designed protein. For example, Wang’s model \citep{wang2022scaffolding}, SMCDiff \citep{trippe2022diffusion} and Tischer's model \citep{tischer2020design} design the scaffold for the specified functional sites. ProteinSGM \citep{lee2022proteinsgm} mask short spans ($<8$ residues) of different secondary structures in different structures and treats the design task as a inpainting problem. CoordVAE \citep{lai2022end} produces novel protein structures conditioned on the backbone template. RefineGNN \citep{jin2021iterative}, CEM \citep{fu2022antibody}, and DiffAb \citep{luoantigen} aim to generate the complementarity-determining regions of the antibody. We summarized protein design model in Table.\ref{tab: protein_design}.

\vspace{-3mm}
\paragraph{Language Diffusion for Protein Structure Generation} Diffusion models \cite{sohl2015deep, ho2020denoising, cao2022survey} are a class of generative models that have achieved impressive results in image \cite{song2020score, lugmayr2022repaint, whang2022deblurring,baranchuk2021label,wolleb2022diffusion}, speech \cite{lee2021nu,chen2020wavegrad,kong2020diffwave,liu2022diffsinger} and text \cite{li2022diffusion,chen2022analog,austin2021structured} synthesis. Recently, FoldingDiff \cite{wu2022protein} shows that language models could be used for unconditional protein generation.

\vspace{-3mm}
\section{Background and Knowledge Gap}
\begin{figure*}[h]
    \centering
    \includegraphics[width=6in]{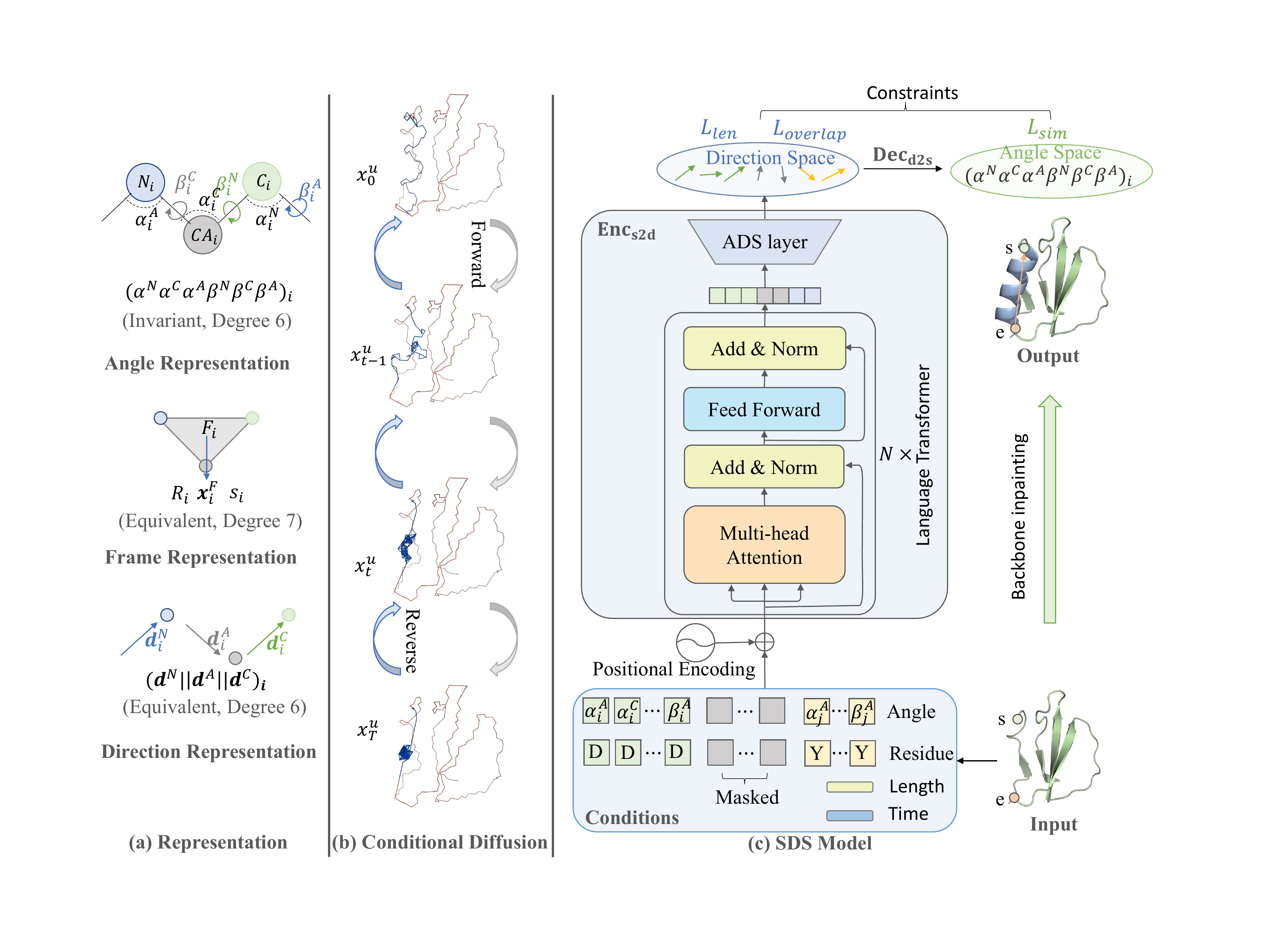}
    \vspace{-3mm}
    \captionsetup{font=small}
    \caption{ The overall framework. (a) We introduce a new direction representation for efficiently computing geometric features while enjoying minimal freedom. (b) Only the masked structures are involved in the diffusion process, while the unmasked ones remain fixed. (c) We add an ADS layer upon the language transformer to convert invariant features into equivalent directions, namely $\text{Enc}_{\text{s2d}}$. Then the $\text{Dec}_{\text{d2s}}$ reverses directions as invariant features based on Eq.\ref{eq:direct2angle}. The geometric constraints ($\mathcal{L}_{len}$ and $\mathcal{L}_{overlap}$) could be efficiently imposed on the direction space. The overall  $\text{Enc}_{\text{s2d}}+\text{Dec}_{\text{d2s}}$ is a simple language model.  }
    \vspace{-3mm}
    \label{fig:framework}
\end{figure*}

Considering three backbone atoms ($N, C_{\alpha}, C$) for each residue, there are 9 ($=3\times 3$) freedom degrees are required for 3D representation.  Recently, researchers have introduced human knowledge into protein backbone representation and proposed two simplified approaches: frame-based and angle-based representation, as shown in Fig.\ref{fig:framework} (a).

\vspace{-3mm}
\paragraph{Frame-based}  This approach treats residues as fundamental elements and assumes that residues of the same type have the same rigid structure, called the local frame. As shown in Fig.\ref{fig:framework} (a), we write $F_i = \{ C_i, C_{\alpha_i}, N_i \}$ as the local frame of the $i$-th residue, where position $\boldsymbol{p}_{F_i}$, orientation $R_i$ and the residue type $s_i$ are required for describing $F_i$, resulting in 7($=3+3+1$) freedom degrees. Under this representation, geometric features can be computed efficiently, e.g., pairwise distance and relative positions. However, the model needs to consider the geometric equivariance of the input data, which introduces considerable modeling complexity.

\vspace{-2mm}
\paragraph{Angle-based} This approach converts structures into sequences of backbone angles based on the order of the protein's primary structure; see Fig.\ref{fig:framework} (a). By assuming the backbone bond lengths are fixed, three bond angles $\alpha^N_i, \alpha^A_i, \alpha^C_i$ and three torsion angles $\beta^N_i, \beta^A_i, \beta^C_i$ are required for describing one residue, leading to 6 freedom degrees. The reduced freedom forms a more compact representation than the frame-based approach. In addition, there is no need to consider geometric equivariance since all angles are invariant to spatial rotation and translation.

\vspace{-2mm}
\paragraph{Knowledge Gap} The angle-based representation seems attractive for simplifying structural modeling and learning more compact protein representations. However, it suffers from the drawback of inefficient computing of geometric features, making it challenging to consider geometric constraints. For example, if one wants to optimize 

\vspace{-5mm}
\begin{align}
    \label{eq:ex_constraint}
    \mathcal{L}_{dist} = \min_{\{\alpha^N_i, \alpha^A_i, \alpha^C_i, \beta^N_i, \beta^A_i, \beta^C_i\}_{0}^i}(||\boldsymbol{p}^A_i - \boldsymbol{p}^A_1|| - r)^2
\end{align}
\vspace{-4mm}

given $\boldsymbol{p}^N_{1}, \boldsymbol{p}^A_{1}, \boldsymbol{p}^C_{1}$. Then, $\boldsymbol{p}^N_{2} \rightarrow \boldsymbol{p}^A_{2}  \rightarrow \boldsymbol{p}^C_{2} \rightarrow \boldsymbol{p}^N_{3} \rightarrow \boldsymbol{p}^A_{3} \rightarrow \cdots \rightarrow \boldsymbol{p}^A_{i}$ needs to be recursively computed by

\vspace{-4mm}
\begin{equation}
    \begin{cases}
        \boldsymbol{p}^{N}_{i}=\text{Place}(\boldsymbol{p}^{C}_{i-1}, \alpha^N_{i-1}, \beta^N_{i-1}, \boldsymbol{d}^C_{i-1}, \boldsymbol{d}^A_{i-1}, r^N)\\
        \boldsymbol{p}^{A}_{i}=\text{Place}(\boldsymbol{p}^{N}_{i}, \alpha^A_i, \beta^A_{i-1}, \boldsymbol{d}^N_i, \boldsymbol{d}^C_{i-1}, r^A)\\
        \boldsymbol{p}^{C}_{i}=\text{Place}(\boldsymbol{p}^{A}_{i}, \alpha^C_{i}, \beta^C_{i}, \boldsymbol{d}^A_{i}, \boldsymbol{d}^N_{i}, r^C)
    \end{cases}
    \label{eq:recursive_place}
\end{equation}
\vspace{-4mm}

where $\boldsymbol{d}^N_i = \frac{\boldsymbol{p}^N_i - \boldsymbol{p}^C_{i-1}}{||\boldsymbol{p}^N_i - \boldsymbol{p}^C_{i-1}||}, \boldsymbol{d}^A_i = \frac{\boldsymbol{p}^A_i - \boldsymbol{p}^N_{i}}{||\boldsymbol{p}^A_i - \boldsymbol{p}^N_{i}||}, \boldsymbol{d}^C_i = \frac{\boldsymbol{p}^C_i - \boldsymbol{p}^A_{i}}{||\boldsymbol{p}^C_i - \boldsymbol{p}^A_{i}||}$. Note that backbone bond lengths, e.g., $r^N = ||\boldsymbol{p}^N_i - \boldsymbol{p}^C_{i-1}||, r^A = ||\boldsymbol{p}^A_i - \boldsymbol{p}^N_{i}||, r^C = ||\boldsymbol{p}^C_i - \boldsymbol{p}^A_{i}||$, are constants. Alg.\ref{alg:place} (in the appendix) shows the details of  $\text{place}(\boldsymbol{p}, \alpha, \beta, \boldsymbol{d}_1, \boldsymbol{d}_2)$. Such recursive computation is inefficient, especially for diffusion models, where the training ($>$15 min/epoch) and generating ($>$1h) overhead is unacceptable. Considering the huge computing overhead, the language model is no longer a simple solution for protein structure design under geometric constraints.

\vspace{-3mm}
\section{Method}
\subsection{Overall Framework}
We propose DiffSDS, inserting direction space into the language diffusion model, to serve as a simple solution for constrained protein backbone inpainting. Specifically, the model takes unmasked atoms and controllable conditions as input to recover the masked region, as shown in Fig.\ref{fig:framework}. Compared to previous backbone inpainting methods, our innovations include the following:

\begin{enumerate}
    \vspace{-2mm}
    \item Inserting direction space into the language model to for simplying the modeling of geometric constraints.
    \vspace{-2mm}
    \item Proposing the first language diffusion model for backbone inpainting and a new modeling perspective.
    \vspace{-2mm}
    \item Introducing geometric conditions and constraints for protein inpainting.
    \vspace{-2mm}
    \item Refining the evaluation metrics, based on which we benchmarked the proposed method and baselines.
    \vspace{-2mm}
 \end{enumerate}

\vspace{-3mm}
\subsection{Direction-based Backbone Representation}
\vspace{-2mm}
\paragraph{ Direction-based Representation}
Is there an alternative representation beyond frame- and angle-based ones to support efficient computation of geometric features while enjoying low degrees of freedom? As shown in Fig.\ref{fig:framework}(a), we introduce direction vectors, i.e., $\boldsymbol{d}^A_i, \boldsymbol{d}^N_i, \boldsymbol{d}^C_i$, for discribing protein structures. In the direction-based representation, the position of each atom is determined by its relative direction ($\boldsymbol{d}^A_i, \boldsymbol{d}^N_i, \boldsymbol{d}^C_i$) and distance ($r^N, r^A, r^C$) from its parent node. In Fig.\ref{fig:framework}(a), taking $C_{\alpha}$ as an example, 

\vspace{-3mm}
\begin{align}
    \label{eq:direct2dist}
    \boldsymbol{p}^A_i =  \boldsymbol{p}^N_i +  r^A \boldsymbol{d}^A_i
\end{align}
\vspace{-3mm}

Recall that $\boldsymbol{p}^A_i$ and $\boldsymbol{p}^N_i$ are spatial coordinates of $C_{\alpha_i}$ and $N_i$.  The direction vector $\boldsymbol{d}^A_i$ points from $\boldsymbol{p}^N_i$ to $\boldsymbol{p}^A_i$, and $r^A$ is the length of the $C_{\alpha}-C$ bond.

\paragraph{Advantages} The propsed representation has several advantages. Firstly, the computing cost of relative positions will be greatly reduced. For example, when computing $\boldsymbol{p}^A_i-\boldsymbol{p}^A_0$, only parallel linear additions and multiplications are required, without recursive computation as in Eq.\ref{eq:recursive_place}:

\vspace{-4mm}
\begin{align}
    \boldsymbol{p}^A_i -  \boldsymbol{p}^A_0=  r^A \sum_{k=1}^i \boldsymbol{d}^A_i + r^C \sum_{k=1}^{i-1} \boldsymbol{d}^C_i + r^N \sum_{k=2}^i \boldsymbol{d}^N_i
    \label{eq:Direct2Pos}
\end{align}
\vspace{-4mm}

Secondly, this representation enjoys the lowest 6 freedom degrees since $||\boldsymbol{d}^A_i|| = ||\boldsymbol{d}^C_i|| = ||\boldsymbol{d}^N_i||=1$. The direction representation could be equivalently transformed to the angle-based one:

\vspace{-3mm}
\begin{equation}
    \begin{cases}
       \alpha^N_i = \arccos(-(\boldsymbol{d}^N_{i+1})^T \boldsymbol{d}^C_{i})\\
       \alpha^A_i = \arccos(-(\boldsymbol{d}^A_{i})^T \boldsymbol{d}^N_{i})\\
       \alpha^C_i = \arccos(-(\boldsymbol{d}^C_{i})^T \boldsymbol{d}^A_{i})\\
       \beta^N_i = \text{dihedral}(\boldsymbol{d}^A_{i}, \boldsymbol{d}^C_{i}, \boldsymbol{d}^N_{i+1})\\
       \beta^A_i = \text{dihedral}(\boldsymbol{d}^C_{i}, \boldsymbol{d}^N_{i+1}, \boldsymbol{d}^A_{i+1})\\
       \beta^C_i = \text{dihedral}(\boldsymbol{d}^N_{i}, \boldsymbol{d}^A_{i}, \boldsymbol{d}^C_{i})
    \end{cases}
    \label{eq:direct2angle}
\end{equation}
\vspace{-3mm}

where $\text{dihedral}(\boldsymbol{v}_1, \boldsymbol{v}_2, \boldsymbol{v}_3)$ is defined in Alg.\ref{alg:dihedral} (Appendix).

 \begin{figure*}[t]
    \centering
    \includegraphics[width=6.5in]{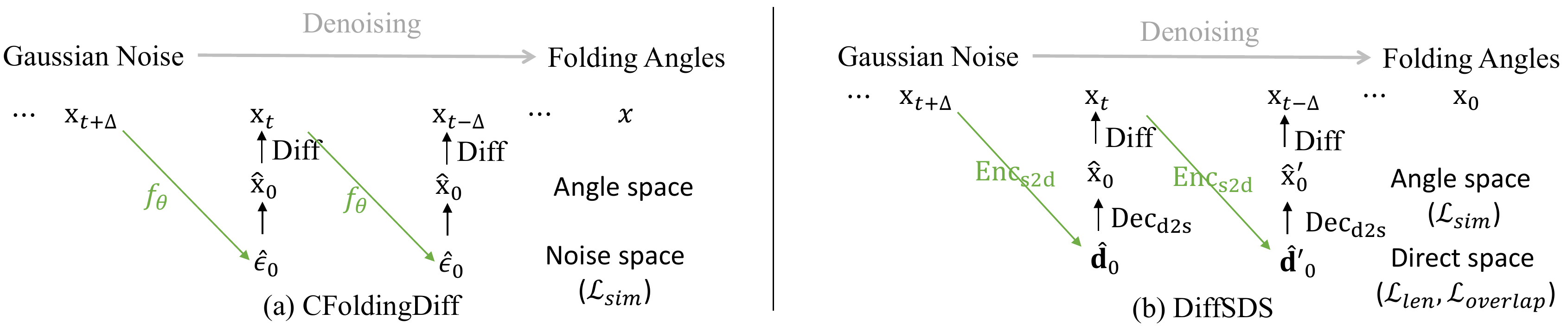}
    \caption{Reverse diffusion. (a) FoldingDiff uses the neural model $f_{\theta}$ to predict one-step noise, where angular similarity loss is imposed on the noise space. We extend FoldingDiff to conditional scenes by fixing the unmasked structure during diffusion to obtain CFoldingDiff. (b) DiffSDS uses $\text{Enc}_{\text{s2d}}$ to predict the direction representation of the original protein. The angular similarity loss and geometric constraints are imposed in the angle and direction space, respectively. }
    \vspace{-4mm}
    \label{fig:diffprocess}
\end{figure*}

\subsection{Conditional language diffusion model equipped with direction space} 
We propose a language diffusion model $f_{\theta}$ equipped with direction space for recovering  masked backbone $\mathcal{M}$ conditional on the unmasked part $\mathcal{U}$:

\vspace{-5mm}
\begin{align}
    \hat{\mathcal{M}} = f_{\theta}(\boldsymbol{p}| \mathcal{U}, x), x\sim \mathcal{N}(0,\mathbf{I})
\end{align}
\vspace{-5mm}

\paragraph{Conditions} The model takes multiple prior conditions as input features for each residue, as shown in Fig.\ref{fig:framework}(c). The node features include backbone angles $F_{ang} \in \mathbb{R}^{n, 6}$, and residue type embedding $F_{res} \in \mathbb{R}^{n, 20}$.  In addition, the global controllable features include the length ($F_{len} \in \mathbb{R}$) of the masked fragment and diffusion timestamp ($F_{time} \in \mathbb{R}$).

\paragraph{Conditional Forward Diffusion} The forward diffusion process could be viewed as a mixup path from clean data $x_0 \sim p_{data}$ to noise $x_T \sim \mathcal{N}(0, I)$:  $x_0 \rightarrow x_{1} \rightarrow \cdots \rightarrow x_T$. Different from generating proteins from scratch, the structure $\mathcal{U}$ is given as a prior, whose angles should not be changed during the diffusion process. Therefore, we divide the latent variable $x_t$ into two parts: $x_t = x_t^{m} \bigoplus x_t^{u}$, where $x_t^m$ and $x_t^u$ are the masked and unmasked protein angles at timestamp $t$. Denote $\alpha_{0}=1, \sigma_{0}=0, q(x_0| x_0) = \mathcal{N}(x_0; \alpha_{0} x, \sigma_{0}^2 I)$, we have

\vspace{-3mm}
\begin{equation}
    \begin{aligned}
        q(x_t^m|x_0^m) &= \mathcal{N}_{\text{wrapped}}(x_t^m; \alpha_t x_0^m, \sigma^2_t I)\\
                    &\propto \sum_{k = -\infty}^{\infty} \exp \left( \frac{-||x_t^m - \alpha_t x_0^m + 2 \pi k ||}{2 \sigma_t^2} \right) 
    \end{aligned}
\end{equation}
\vspace{-3mm}

where we use the wrapped normal \citep{wu2022protein} to force the angles space in $[0, \pi]$. The hyper-parameters $\alpha_{t}$ and $\sigma_{t}$ determine the diffusion schedule:

\begin{equation}
    \begin{cases}
      \sigma_{t} = \text{clip}(1-\alpha_{t}, 0.999)\\
      \alpha_{t} = \cos{(t/T \cdot \frac{\pi}{2})}
    \end{cases}
\end{equation}

\paragraph{Direction-aware Reverse Diffusion} The reverse process applies neural network $f_{\theta}$ as the translation kernel to recover clean data following the Markov chain $x_T \rightarrow x_{T-1} \rightarrow \cdots \rightarrow x_0$. As derived in the Appendix, the objective is to maximize $q(x_{t-1}| x_{t}, x_0) = \mathcal{N}(z_{t-1}; \hat{\mu}_{t-1}, \hat{\sigma}_{t-1}^2 I)$, and 

\vspace{-3mm}
\begin{equation}
 \begin{cases}
   \hat{\sigma}_s = \frac{\sigma_{t|s}\sigma_{s}}{\sigma_{t}}\\
   \hat{\mu}_s = \frac{1}{\alpha_{t|s}} x_t - \frac{\alpha_{t|s}}{\sigma_t} \epsilon_t
 \end{cases}
\end{equation}
\vspace{-3mm}

There are several alternative variables could be estimated by $f_{\theta}$ to obtain $q(x_{t-1}| x_{t}, x_0)$, such as $\hat{\mu}_{t-1}, \hat{\epsilon}_{t-1}, \hat{x}_0$. As shown in Fig.\ref{fig:diffprocess}, FoldingDiff realizes the neural network as $f_{\theta}: (x_t, t) \mapsto \hat{\epsilon}_t$, which is effective for unconditional protein backbone generation. Instead, we prefer $f_{\theta}: (x_t, t) \mapsto \hat{x}_0$ and decompose it as encoder $\text{Enc}_{\text{s2d}}$ and decoder $\text{Dec}_{\text{d2s}}$. The $\text{Enc}_{\text{s2d}}: (x_t, t) \mapsto \hat{d}_0$ predicts the direction vectors (${d}_0$) of the backbone, and the decoder $\text{Dec}_{\text{d2s}}: \hat{d}_0 \mapsto \hat{x}_0$ reverses the direction representation into angle representation based on Eq.\ref{eq:direct2angle}. With the inserted direction space, we could efficiently compute geometric features from and impose corresponding constraints on the model, as illustrated in Eq.\ref{eq:direct2dist} and Eq.\ref{eq:ex_constraint}. More importantly, this modification does not increase the modeling complexity: $f_{\theta}$ still appears as a language model, with the inputs and outputs being sequences.

\vspace{-3mm}
\subsection{Constraints}
\paragraph{Protein likeness} From the diffusion perspective, the neural model needs to recover the masked backbone angles to ensure the protein likeness. The overall objective is to maximize the variational lower bound of $\log{p_{\theta}(x_0)}$:

\vspace{-3mm}
\begin{equation*}
    \resizebox{0.48 \textwidth}{!}{$
    \begin{split}
        &\mathcal{L}_{vlb}(x_0) = \\ &\mathop{\mathbb{E}}_{q(x_{1:T} | x_{0})} \left[\log \frac{q(x_{T} | x_{0})}{\ptheta(x_{T})} + \sum_{t=2}^T \log \frac{q(x_{t-1} | x_{0},x_{t})} {\ptheta(x_{t-1} | x_{t}) } - \log \ptheta( x_{0} | x_{1})\right]
    \end{split}
    $}
\end{equation*}
\vspace{-3mm}

In practice, we use the simplfied loss:

\vspace{-3mm}
\begin{equation*}
    \begin{split}
        &\mathcal{L}_{sim}(x_0) = \sum_{t=0}^{T} ||f_{\theta}(x_t,t) - x_0||^2
    \end{split}
\end{equation*}
\vspace{-3mm}

\paragraph{Length loss} To ensure the designed $\hat{\mathcal{M}}$ has a similar length as the reference structure $\mathcal{M}$, such that the masked endpoints ($s$ and $e$) could be  connected, we impose the length loss on the hidden direction space:

\vspace{-3mm}
\begin{align}
    \mathcal{L}_{len} =& \sum_{S \in \{N,A,C\}} (||\hat{\boldsymbol{p}}^S_s - \hat{\boldsymbol{p}}^S_e|| - ||\boldsymbol{p}^S_s - \boldsymbol{p}^S_e||)^2
\end{align}
\vspace{-3mm}

where $\hat{\boldsymbol{p}}^S_s - \hat{\boldsymbol{p}}^S_e$ are computed by Eq.\ref{eq:Direct2Pos} using the output directions of $\text{Enc}_{s2d}$.

\paragraph{Overlapping loss} To avoid overlapping between designed $\hat{\mathcal{M}}$ and unmasked structure $\mathcal{U}$, the overlapping loss is also imposed on the direction space:

\begin{align}
    \mathcal{L}_{overlap} =& \sum_{i \in 
    [s+1,e-1]} e^{-\tau ||\hat{\boldsymbol{p}}^A_i - \boldsymbol{p}^A_j||}
\end{align}

where $j = \min_{j \in [1,s] \cup [e,n]} || \boldsymbol{p}^A_i - \boldsymbol{p}^A_j ||$ is the nearst $\alpha$-carbon atom to $\hat{\boldsymbol{p}}^A_i$ in $\mathcal{U}$. By default, we set $\tau=0.8$.

\paragraph{Overall loss} During training, we impose protein similarity loss ($\mathcal{L}_{sim}$), length loss ($\mathcal{L}_{len}$), and overlapping loss ($\mathcal{L}_{overlap}$) on the model, the overall loss function is:

\vspace{-3mm}
\begin{align}
    \mathcal{L} = \lambda_1 \mathcal{L}_{sim} + \lambda_2 \mathcal{L}_{len} + \lambda_3 \mathcal{L}_{overlap}
\end{align}

where we choose $\alpha_1 = 1, \alpha_2 = 0.001, \alpha_3=10$.

\section{Experiments}
In this section, we conduct extensive experiments to evaluate the proposed method. Specifically, we would like to answer the following questions:

\begin{itemize}%[leftmargin=*]
    \vspace{-2mm}
    \item \textbf{Q1: Comparision} Do the structures generated by DiffSDS achieve better protein likeness, connectivity, and non-overlapping metrics than baselines?
    \vspace{-2mm}
    \item \textbf{Q1: Ablation} How can conditional features and constraints improve the model performance?
    \vspace{-2mm}
    \item \textbf{Q3: Designability} Are the generated structures likely to be designed?
    \vspace{-2mm}
\end{itemize}

\subsection{Overall Setting}
\paragraph{Data split} We train models on CATH4.3, where proteins are partitioned by the CATH topology classification. To avoid potential information leakage, we further refine the test set by excluding proteins that are similar to the training data from the test set, i.e., TM-score greater than 0.5. Finally, there are 24,199 proteins for training, 3,094 proteins for validation, and 378 proteins for testing. 

\paragraph{Baselines \& Training Setting}  We compare DiffSDS with the recent strong baselines RFDesign \cite{wang2022scaffolding} and CFoldingDiff.  RFDesign is the state-of-the-art model trained across the whole PDB dataset and accepted by the Science journal. CFoldingDiff is a derivative of FoldingDiff \cite{wu2022protein} where the angles of the unmasked residues are fixed during diffusion. We evaluate RFDesign, CFoldingDiff, and DiffSDS on the same test dataset, where the contiguous backbone of length $m \sim U(5,L/3)$ were randomly masked, given the protein length  $L$. For the same protein, the masked area keeps the same when evaluating different methods. We retrain CFoldingDiff to make it suitable for the inpainting task, while the pre-trained RFDesign model can be used directly. As to DiffSDS, we use 16 transformer layers, with 384 hidden dimensions and 12 attention heads per layer. We train CFoldingDiff and DiffSDS up to 10000 epochs with an early stop patience of 1000, where the learning rate is 0.0001, the batch size is 128, and the maximum diffusion timestamp $T$ is 1000. All experiments are conducted on NVIDIA-A100s

\subsection{Protein likeness}
\begin{figure*}[t]
    \centering
    \includegraphics[width=6.8in]{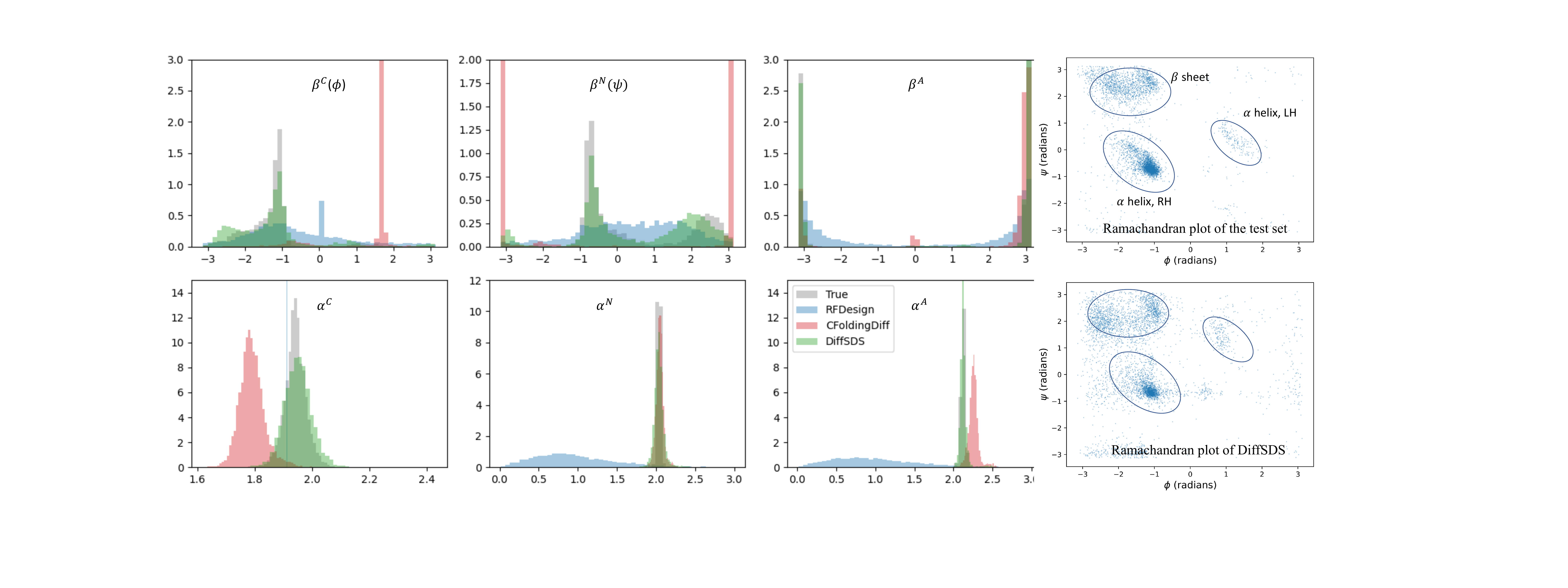}
    \vspace{-4mm}
    \caption{ Angle distributions of various methods vs the test set distribution. DiffSDS produces the most similar angle distributions to those of the test set. Ramachandran plots also show that DiffSDS can produce realistic structural distributions}
    \vspace{-4mm}
    \label{fig:angles}
  \end{figure*}

\paragraph{Objective \& Setting} Are the designed structures likely to constitute native proteins? We take Rosetta energies (rama and omega) as metrics to measure the protein likeness of the generated backbones. The "rama" indicates Ramachandran torsion energy derived from statistics on the PDB, and "omega" indicates omega angle energy. In addition, we show the angle distributions and Ramachandran plots of the different methods in Fig.\ref{fig:angles}. We group results by the masked length to reveal the performance at different mask lengths. We also adopt the energy of the test set structures as a baseline to show how closely we approximate the reference structures.

% 角度分布图

\begin{table}[h]
    \resizebox{1.0 \columnwidth}{!}{
    \begin{tabular}{ccccccc}
        \toprule
    energy       & \multicolumn{3}{c}{rama} & \multicolumn{3}{c}{omega} \\ \cline{2-4} \cline{5-7}
    mask length  & $<$15    & 15-30   & $>$30   & $<$15    & 15-30    & $>$30   \\ \hline
    Test         &  0.67      &   0.71      &  0.62     &  0.75      &   0.68       &   0.66  \\
    RFDesign     &  {2.12}      &   {2.49}      &  {3.38}     &  {16.62}      &   {12.56}       &   {13.14}    \\
    CFoldingDiff  &  \underline{1.65}      &   \underline{1.86}      &  \underline{2.11}   &  \underline{6.26}      &    \underline{4.97}      &  \underline{4.17}     \\
    Dualspace &  \textbf{1.51}      &   \textbf{1.76}      &  \textbf{1.95}     &  \textbf{4.30}      &    \textbf{3.17}      &  \textbf{2.77}  \\  
    \bottomrule 
\end{tabular}}

\vspace{-3mm}
\caption{Rosetta energies of generated backbones. The \textbf{best} and \underline{suboptimal} results are labeled with bold and underline.}
\label{tab:energy}
\end{table}

\vspace{-3mm}
\paragraph{Results \& Analysis} As shown in Table.\ref{tab:energy}, DiffSDS achieves the best scores on both 'omega' and 'rama' energies, indicating that the structure generated by DiffSDS is more likely to be a native structure. These improvements are consistent in terms of different masked lengths. However, there is still a large performance gap compared to the test set energies, which suggests that there is still a long way to go in designing protein backbones. In Figure.\ref{fig:angles}, we compare the angular distributions of different methods, where DiffSDS's results are the closest to the test set distribution. 

\begin{figure*}[b]
    \centering
    \includegraphics[width=5.6in]{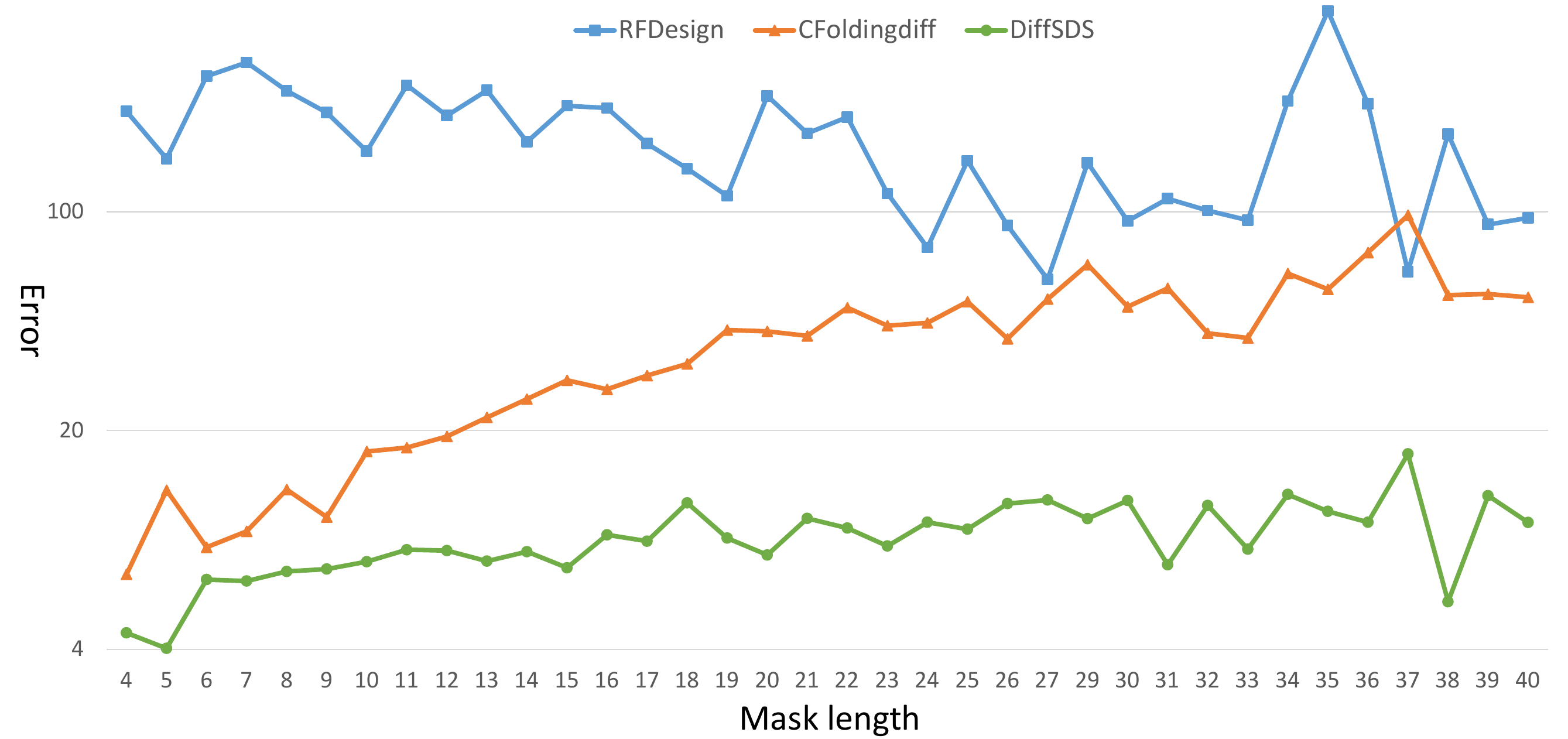}
    \vspace{-4mm}
    \caption{The connectivity error trend of different methods. }
    \label{fig:connectivity}
\end{figure*}

\subsection{Connectivity}
\paragraph{Objective \& Setting} Do the designed structures connect the endpoints without breakage? As shown in Figure.\ref{fig:break}, the connectivity is an important indicator for the inpainting task, as disconnected proteins must be structurally abnormal. However, this metric has lacked  attention in previous work, and we begin by defining the connectivity error as:

\vspace{-3mm}
\begin{align}
    \text{Error} = ||\boldsymbol{p}^A_{s} - \boldsymbol{\hat{p}}^A_{s}|| + ||\boldsymbol{p}^A_{e} - \boldsymbol{\hat{p}}^A_{e}||
\end{align}
\vspace{-3mm}

where $s$ and $e$ are indexes of the start and end points of the masked structure, $\boldsymbol{\hat{p}}^A$ and $\boldsymbol{p}^A$ indicate the predicted and ground truth positions of the $C_{\alpha}$-carbon. Ideally, $\text{Error}=0$ means that endpoints are connected as expected.

\paragraph{Results \& Analysis} From Table.\ref{tab:connectivity_error}, we conclude that DiffSDS could achieve the lowest connectivity error compared to RFDesign and FoldingDiff, suggesting that it can better connect masked endpoints. We further show the trend of connectivity error with increasing mask length in Figure.\ref{fig:connectivity}, from which we find that RFDesign performs poorly at all mask lengths, FoldingDiff's connectivity error increases with mask length, while DiffSDS performs steadily and consistently better than all baselines. 

\vspace{-3mm}
\begin{table}[h]
    \centering
    \begin{tabular}{cccc}
    \toprule
               & \multicolumn{3}{c}{connectivity error}\\ 
    mask length  & $<$10    & 10-15   & $>$15     \\\hline
    % Test set  & 0 & 0 & 0 \\
    RFDesign     &  218.36      &   146.24      &   159.17        \\
    FoldingDiff  &  \underline{14.77}      &  \underline{42.65}       &   \underline{59.08}        \\
    DiffSDS &  \textbf{6.93}      &   \textbf{9.93}      &    \textbf{10.61}       \\ 
    \bottomrule
    \end{tabular}
    \vspace{-3mm}
    \caption{connectivity error of different methods.}
    \label{tab:connectivity_error}

\end{table}

\vspace{-5mm}
\begin{figure*}[b]
    \centering
    \includegraphics[width=5.7in]{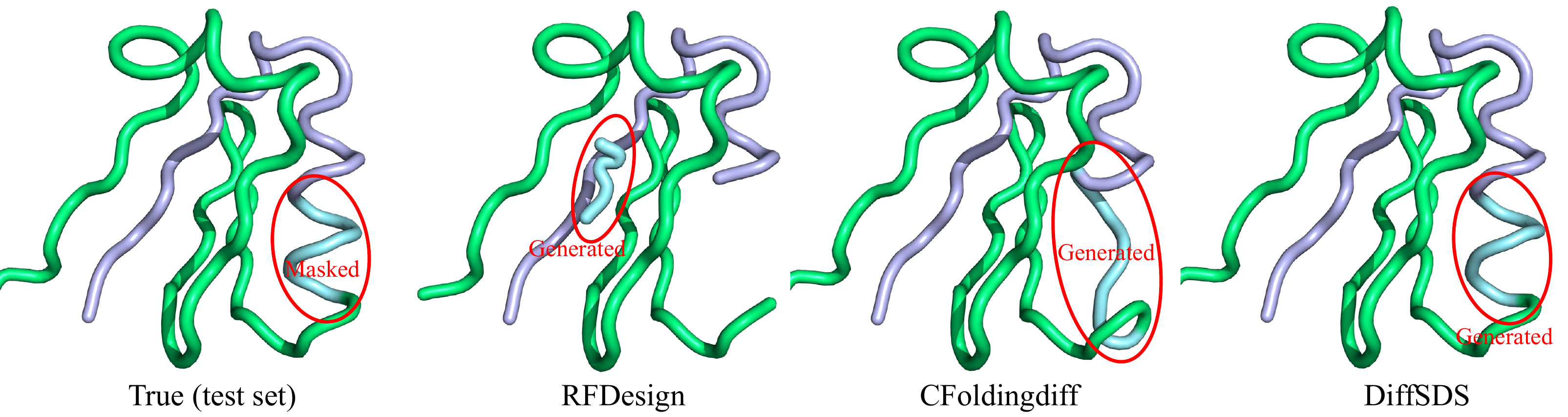}
    \vspace{-4mm}
    \caption{A protein inpating example. DiffSDS generates the most similar structure to the reference protein. }
    \label{fig:design_example}
\end{figure*}
\vspace{-5mm}

\subsection{Non-overlapping}

% 计算overlap
\paragraph{Objective \& Setting} Will the designed structures be overlapped with existing backbones? We evaluate the spatial interaction between the generated structure ($\hat{\mathcal{M}}$) and the unmasked structure ($\mathcal{U}$), and define the interaction score as 

\vspace{-4mm}
\begin{align}
    \text{Score}_d = \sum_{i \in \hat{\mathcal{M}}, j \in \mathcal{U}} \mathds{1}(||\boldsymbol{\hat{x}}_i - \boldsymbol{\hat{x}}_j||<d)
\end{align}
\vspace{-4mm}

where $\mathds{1}(\cdot)$ is an indicator function. $\text{Score}_d$ records the number of pairwise interactions between masked and non-masked amino acids, with distances threshold $d$ $\mathring{A}$.

\begin{table}[h]
    \centering
    \resizebox{1.0 \columnwidth}{!}{
    \begin{tabular}{ccccccc}
        \toprule
                & \multicolumn{3}{c}{$d=1\mathring{A}$}                             & \multicolumn{3}{c}{$d=3\mathring{A}$} \\ \cline{2-4} \cline{5-7}
    mask length & $<$15           & 15-30           & $>$30           & $<$15  & 15-30 & $>$30 \\
    \hline
    True        & 266             & 345             & 139             & 268    & 406   & 139   \\
    RFDesign    & 280             & 357             & 153             & 861    & 1362  & 593   \\
    Foldingdiff & \underline{276} & \textbf{352}    & \underline{145} & \underline{552}    & \underline{640}   & \underline{222}   \\
    DiffSDS   & \textbf{270}    & \underline{356} & \textbf{141}    & \textbf{472}    & \textbf{632}   & \textbf{178}  \\
    \bottomrule
    \end{tabular}}
    \vspace{-3mm}
    \caption{The number of spatial interactions.}
    \vspace{-3mm}
    \label{tab:overlap}
\end{table}

\vspace{-3mm}
\paragraph{Results \& Analysis} As shown in Table.\ref{tab:overlap}, the spatial interactions of DiffSDS's generated structures are closer to the test set than the baselines. This verifies that the non-overlap loss we introduced can avoid spatial overlap.

\vspace{-3mm}
\subsection{Ablation study}
\paragraph{Objective\&Setting} We conduct ablation experiments to investigate the impact of conditions and constraints on protein backbone inpainting. Specifically, we show how these factors affect the validation losses, including angle loss, length loss, and overlapping loss.

\vspace{-3mm}
\paragraph{Results \& Analysis} Limited to the page, we show the ablation details in the Appendix. As shown in Fig.\ref{fig:ablation_cond} and Fig.\ref{fig:ablation_cons}, we find that: (1) the length condition and the sequence condition contribute to the reduction of $\mathcal{L}_{len}$ and $\mathcal{L}_{sim}$, respectively. This phenomenon suggests that the model can learn to generate structures with a predetermined length and that residue types can facilitate learning backbone angles. (2) Explicitly imposing geometric constraints on the model is necessary. If the constraints are removed, the $\mathcal{L}_{len}$ loss is difficult to reduce, and the $\mathcal{L}_{overlap}$ even increases, indicating the model could not generate geometrically reasonable structures. Fortunately, all these drawbacks could be eliminated by imposing geometric loss on the introduced direction space.

\vspace{-3mm}
\subsection{Designability}
\paragraph{Objective \& Setting} How likely are the generated proteins to be synthesized in the laboratory? We further measure the designability of the generated protein by the self-consistency TM score (scTM), which is first introduced by  FoldingDiff. The generated backbones are fed into ESM-IF to obtain candidate protein sequences, which are subsequently folded into 3D structures by OmegaFold. scTM is the TM score between the newly folded structure and the original generated structure. If scTM$>0.5$, the corresponding generated backbone is considered designable.

\vspace{-3mm}
\paragraph{Results \& Analysis} As shown in Table\ref{tab:Designability}, the designability order is: DiffSDS$>$CFoldingDiff$>$RFDesign. On the full test set, DiffSDS generate 231 designable backbones, outperforming CFoldingDiff's 217 and RFDesign's 178. For proteins of more than 70 residues, the relative improvement is 8.6\%; for long proteins of more than 70 residues, the relative improvement achieves 10\%. Besides,  DiffSDS outperforms previous methods in both average and median scTM scores. In Fig.\ref{fig:design_example}, we show a protein inpainting example, comparing different methods.

\vspace{-3mm}
\begin{table}[h]
    \resizebox{1.0 \columnwidth}{!}{
    \begin{tabular}{cccccc}
        \toprule
                 & \multicolumn{3}{c}{scTM$>$0.5} & \multicolumn{2}{c}{scTM} \\
                 & All      & len$\leq$70 & len$>$70  & Mean       & Median      \\\midrule
    RFDesign     & 178/378  & 65/148  & 113/230 & 0.51       & 0.48        \\
    CFoldingDiff & \underline{217/378}  & \underline{81/148}  & \underline{130/230} & \underline{0.54}       & \underline{0.53}        \\
    DiffSDS      & \textbf{231/378}  & \textbf{88/148}  & \textbf{143/230} & \textbf{0.56}       & \textbf{0.55}       \\
    Improvement & 6.5\% & 8.6\% & 10\% & 3.7\% & 3.8\%\\
    \bottomrule
    \end{tabular}}
    \vspace{-3mm}
    \caption{Designability of different methods.}
    \label{tab:Designability}
    \vspace{-3mm}
\end{table}

\vspace{-3mm}
\section{Conclusion}
This paper introduces and inserts a direction-based representation space on the language model to support efficient geometric feature computation and maintain modeling simplicity. By imposing geometric constraints on the direction space and applying the model during conditional diffusion, the proposed DiffSDS achieves significant improvements in protein backbone inpainting compared to baselines.

\bibliography{DiffSDS}
\bibliographystyle{icml2023}

%%%%%%%%%%%%%%%%%%%%%%%%%%%%%%%%%%%%%%%%%%%%%%%%%%%%%%%%%%%%%%%%%%%%%%%%%%%%%%%
%%%%%%%%%%%%%%%%%%%%%%%%%%%%%%%%%%%%%%%%%%%%%%%%%%%%%%%%%%%%%%%%%%%%%%%%%%%%%%%
% APPENDIX
%%%%%%%%%%%%%%%%%%%%%%%%%%%%%%%%%%%%%%%%%%%%%%%%%%%%%%%%%%%%%%%%%%%%%%%%%%%%%%%
%%%%%%%%%%%%%%%%%%%%%%%%%%%%%%%%%%%%%%%%%%%%%%%%%%%%%%%%%%%%%%%%%%%%%%%%%%%%%%%
\newpage

\appendix
\onecolumn

\section{Appendix}
\subsection{Related works}

\paragraph{3D Molecule Generation} Generating 3D molecules to explore the local minima of the energy function (Conformation Generation) \cite{gebauer2019symmetry, 10.5555/3524938.3525768, simm2020reinforcement, shi2021learning, xu2021end, luo2021predicting, xu2020learning, ganea2021geomol, xu2022geodiff, hoogeboom2022equivariant, jing2022torsional, zhu2022direct} or discover potential drug molecules binding to targeted proteins (3D Drug Design)  \cite{imrie2020deep, nesterov20203dmolnet, luo20223d, ragoza2022chemsci, wu2022diffusion, huang2022mdm, peng2022pocket2mol, huang20223dlinker, wang2022generative, liu2022generating} have attracted extensive attention in recent years. Compared to conformation generation that aims to predict the set of favourable conformers from the molecular graph, 3D Drug Design is more challenging in two aspects: (1) both conformation and molecule graph need to be generated, and (2) the generated molecules should satisfy multiple constraints, such as physical prior and protein-ligand binding affinity. We summarized representive works of 3D drug design in Table.\ref{tab: mol_generate_models} in the appendix, where all the methods focus on small molecule design.

\begin{table}[h]
    \centering
    \caption{3D molecule generation models. }
    \label{tab: mol_generate_models}
    
    \resizebox{0.5 \columnwidth}{!}{
    \begin{tabular}{ccc}
      \toprule
    Method      & Input & Github  \\ \midrule
    \multicolumn{3}{c}{Molecule Conformation Generation}\\ \hline
    G-SchNet \cite{gebauer2019symmetry} & -- & \href{https://github.com/atomistic-machine-learning/G-SchNet}{PyTorch} \\
    CVGAE \cite{mansimov2019molecular} & 2D-graph & \href{https://github.com/nyu-dl/dl4chem-geometry}{TF} \\
    GraphDG \cite{10.5555/3524938.3525768} & 2D-graph & \href{https://github.com/gncs/graphdg}{PyTorch} \\
    MolGym \cite{simm2020reinforcement} & -- & \href{https://github.com/gncs/molgym}{PyTorch} \\
    ConfGF \cite{shi2021learning} & 2D-graph & \href{https://github.com/DeepGraphLearning/ConfGF}{PyTorch} \\
    ConfVAE \cite{xu2021end} & 2D-graph & \href{https://github.com/MinkaiXu/ConfVAE-ICML21}{PyTorch} \\
    DGSM \cite{luo2021predicting} & 2D-graph & -- \\
    CGCF \cite{xu2020learning} & 2D-graph & \href{https://github.com/DeepGraphLearning/CGCF-ConfGen}{PyTorch} \\
    GeoMol \cite{ganea2021geomol} & 2D-graph & \href{https://github.com/PattanaikL/GeoMol}{PyTorch} \\
    G-SphereNet \cite{luo2021autoregressive} & -- & \href{https://github.com/divelab/DIG}{PyTorch} \\
    GeoDiff \cite{xu2022geodiff} & 2D-graph & \href{https://github.com/MinkaiXu/GeoDiff}{PyTorch}\\
    EDM \cite{hoogeboom2022equivariant} & 2D-graph & \href{https://github.com/ehoogeboom/e3_diffusion_for_molecules.git}{PyTorch} \\
    TorsionDiff \cite{jing2022torsional} & 2D-graph & \href{https://github.com/gcorso/torsional-diffusion}{PyTorch} \\
    DMCG \cite{zhu2022direct} & 2D-graph & \href{https://github.com/DirectMolecularConfGen/DMCG}{PyTorch} \\ 
    \hline 
     \multicolumn{3}{c}{\textit{De novo} Molecule Design}\\ \hline
    DeLinker \cite{imrie2020deep} & \begin{tabular}[c]{@{}c@{}}Protein Pocket\\ 3D-fragments\end{tabular}  & \href{https://github.com/oxpig/DeLinker}{TF} \\
    3DMolNet \cite{nesterov20203dmolnet} & 3D-geometry & -- \\
    cG-SchNet \cite{gebauer2022inverse} & 3D-geometry & \href{https://github.com/atomistic-machine-learning/cG-SchNet}{PyTorch} \\
    Luo's model \cite{luo20223d} & Protein Pocket & \href{https://github.com/luost26/3D-Generative-SBDD}{PyTorch}\\
    LiGAN \cite{ragoza2022chemsci} & Protein Pocket & \href{https://github.com/mattragoza/LiGAN}{PyTorch} \\
    Bridge \cite{wu2022diffusion} & Physical prior & -- \\
    MDM \cite{huang2022mdm} & \begin{tabular}[c]{@{}c@{}}2D-graph \\ Properties\end{tabular} & -- \\
    Pocket2Mol \cite{peng2022pocket2mol} & Protein Pocket & \href{https://github.com/pengxingang/Pocket2Mol}{PyTorch} \\
    3DLinkcer \cite{huang20223dlinker} & 3D-fragments & \href{https://github.com/YinanHuang/3DLinker}{PyTorch} \\
    CGVAE \cite{wang2022generative} & Coarse Topology & \href{https://github.com/wwang2/coarsegrainingvae}{PyTorch} \\
    GraphBP \cite{liu2022generating} &  Protein Pocket & \href{https://github.com/divelab/GraphBP}{PyTorch} \\ 

    \bottomrule
    \end{tabular}}
\end{table}

\paragraph{Protein Design} In addition to small molecules, biomolecules such as proteins have also attracted considerable attention by researchers \citep{ding2022protein, ovchinnikov2021structure, gao2020deep, strokach2022deep}. We divide the mainstream protein design methods into three categories: protein sequence design \citep{li2014direct, wu2021protein, pearce2021deep, ingraham2019generative, jing2020learning, tan2022generative, gao2022alphadesign, hsu2022learning, dauparas2022robust, gao2022pifold, o2018spin2, wang2018computational, qi2020densecpd, strokach2020fast, chen2019improve, zhang2020prodconn, anand2022protein}, unconditional protein structure generation \citep{anand2018generative, sabban2020ramanet, eguchi2022ig, wu2022protein}, and conditional protein design \citep{lee2022proteinsgm, wang2022scaffolding, trippe2022diffusion, lai2022end, fu2022antibody, tischer2020design, anand2022protein, luoantigen}. Protein sequence design aims to discover protein sequences folding into the desired structure, and unconditional protein structure generation focus on generating new protein structures from noisy inputs. We are interested in conditional protein design and consider multiple constraints on the designed protein. For example, Wang’s model \citep{wang2022scaffolding}, SMCDiff \citep{trippe2022diffusion} and Tischer's model \citep{tischer2020design} design the scaffold for the specified functional sites. ProteinSGM \citep{lee2022proteinsgm} mask short spans ($<8$ residues) of different secondary structures in different structures and treats the design task as a inpainting problem. CoordVAE \citep{lai2022end} produces novel protein structures conditioned on the backbone template. RefineGNN \citep{jin2021iterative}, CEM \citep{fu2022antibody}, and DiffAb \citep{luoantigen} aim to generate the complementarity-determining regions of the antibody. We summarized protein design model in Table.\ref{tab: protein_design}.

\begin{table}[h]
  \centering
  \caption{Protein Design Models. }
  \label{tab: protein_design}
  \resizebox{0.5 \columnwidth}{!}{
  \begin{tabular}{ccc}
    \toprule
  Method      & Input & Github  \\ \midrule
  \multicolumn{3}{c}{Unconditional protein structure generation}\\ \hline
  Anand's model \citep{anand2018generative} & Noise & \href{https://github.com/collinarnett/protein_gan}{PyTorch}\\
  RamaNet \citep{sabban2020ramanet} & Noise & \href{https://github.com/sarisabban/RamaNet}{TF}\\
  Ig-VAE \citep{eguchi2022ig} & Noise & \href{https://github.com/ProteinDesignLab/IgVAE}{PyTorch}\\
  FoldingDiff \citep{wu2022protein} & Noise & \href{https://github.com/microsoft/foldingdiff}{PyTorch}\\\hline

  \multicolumn{3}{c}{Protein seqeunce design}\\ \hline
  GraphTrans \cite{ingraham2019generative}  & 3D Backbone &\href{https://github.com/jingraham/neurips19-graph-protein-design}{PyTorch}\\
  GVP \citep{jing2020learning}  & 3D Backbone & \href{https://github.com/drorlab/gvp-pytorch}{PyTorch}\\
  GCA \citep{tan2022generative} & 3D Backbone & \href{https://github.com/chengtan9907/Global-context-aware-generative-protein-design}{PyTorch}\\
  AlphaDesign \citep{gao2022alphadesign} & 3D Backbone & \href{https://github.com/Westlake-drug-discovery/AlphaDesign}{PyTorch}\\
  ESM-IF \citep{hsu2022learning} & 3D Backbone & \href{https://github.com/facebookresearch/esm/tree/main/examples/inverse_folding}{PyTorch}\\
  ProteinMPNN \citep{dauparas2022robust} & 3D Backbone & \href{https://github.com/dauparas/ProteinMPNN}{PyTorch}\\
  PiFold \citep{gao2022pifold} & 3D Backbone & \href{https://github.com/A4Bio/PiFold}{PyTorch}\\\hline

  \multicolumn{3}{c}{Conditional protein design}\\ \hline
  ProteinSGM \citep{lee2022proteinsgm} & Masked structures & -- \\
  Wang's model \cite{wang2022scaffolding}    & Functional sites & \href{https://github.com/RosettaCommons/RFDesign}{PyTorch} \\
  SMCDiff \citep{trippe2022diffusion} & Functional motifs & --\\
  CoordVAE \cite{lai2022end} & Backbone Template & -- \\
  CEM \cite{fu2022antibody} & CDR geometry & -- \\
  Tischer's model \citep{tischer2020design}  & Functional motifs &\href{https://github.com/dtischer/trdesign-motif}{TF}\\
  Anand's model \citep{anand2022protein} & Multiple conditions & --\\
  RefineGNN \citep{jin2021iterative} & Antigen structure & \href{https://github.com/wengong-jin/RefineGNN}{PyTorch}\\
  DiffAb \citep{luoantigen} & Antigen structure & \href{https://github.com/luost26/diffab}{PyTorch}\\

  \bottomrule
  \end{tabular}}
\end{table}

\subsection{Algorithms}

\begin{algorithm}[h]
  \caption{$\text{place}(\boldsymbol{x}, \alpha, \beta, \boldsymbol{d}_1, \boldsymbol{d}_2, r)$}
  \label{alg:place}
\begin{algorithmic}[1]
  \STATE {\bfseries Input:} $\boldsymbol{x}, \alpha, \beta, \boldsymbol{d}_1, \boldsymbol{d}_2$  
  \STATE $\boldsymbol{\tilde{d}}=[- \cos{\alpha}, \cos{\beta} \sin{\alpha}, \sin{\beta} \sin{\alpha}]^T$
  \STATE $R = [\boldsymbol{d}_1, (\boldsymbol{d}_2 \times \boldsymbol{d}_1) \times \boldsymbol{d}_1, \boldsymbol{d}_2 \times \boldsymbol{d}_1]$
  \STATE $\boldsymbol{d} = R \boldsymbol{\tilde{d}}$
  \STATE {\bfseries Return:} $\boldsymbol{x} + r \boldsymbol{d}$
\end{algorithmic}
\end{algorithm}

\begin{algorithm}[h]
  \caption{$\text{dihedral}(\boldsymbol{v}_1, \boldsymbol{v}_2, \boldsymbol{v}_3)$}
  \label{alg:dihedral}
\begin{algorithmic}[1]
  \STATE {\bfseries Input:} $\boldsymbol{v}_1, \boldsymbol{v}_2, \boldsymbol{v}_3$  
  \STATE $\boldsymbol{n}_1 = \boldsymbol{v}_1 \times \boldsymbol{v}_2$
  \STATE $\boldsymbol{n}_2 = \boldsymbol{v}_2 \times \boldsymbol{v}_3$
  \STATE $x = (\boldsymbol{n}_1)^T \boldsymbol{n}_2$
  \STATE $y = \boldsymbol{n}_1 \times\boldsymbol{n}_2$
  \STATE {\bfseries Return:} $\arctan{\frac{y}{x}}$
\end{algorithmic}
\end{algorithm}

\clearpage
\subsection{Diffusion}

\paragraph{Forward process}
We start from the standard diffusion process $x_0 \rightarrow x_1 \rightarrow \cdots \rightarrow x_T$, where the forward translation kernel from timestamp $s$ to $t$ is defined as $q(x_t | x_{s}) =  \mathcal{N}(x_t; \alpha_{t|s} x_{s}, \sigma_{t|s}^2 I)$, $s \leq t$. Denote $\alpha_t = \alpha_{t|0}, \sigma_t = \sigma_{t|0}$, and $q(x_0| x_0) = \mathcal{N}(x_0; \alpha_{0} x, \sigma_{0}^2 I), \alpha_{0}=1, \sigma_{0}=0$. We will show that $\alpha_{t|s} = \alpha_t / \alpha_s, \sigma^2_{t|s} = \sigma^2_{t} - \alpha^2_{t|s} \sigma^2_{s}$.

\begin{proof}

  \begin{equation}
    \begin{aligned}
       q(x_t | x_{s}) =&  \mathcal{N}(x_t; \alpha_{t|s} x_{s}, \sigma_{t|s}^2 I) \\
       \Rightarrow x_t =& \alpha_{t|t-1} x_{t-1}  + \sigma_{t|t-1} \epsilon_{t-1} & \textcolor{gray}{ x_t \sim  q(x_t | x_{t-1})}\\
        =& \alpha_{t|t-1} ( \alpha_{t-1|s} x_{s}  + \sigma_{t-1|s} \epsilon_{s} ) 
          + \sigma_{t|t-1} \epsilon_{t-1} & \textcolor{gray}{ x_{t-1} \sim  q(x_{t-1} | x_{s})}\\  
        =& \alpha_{t|t-1} \alpha_{t-1|s} x_{s} + \alpha_{t|t-1} \sigma_{t-1|s} \epsilon_{s}  + \sigma_{t|t-1} \epsilon_{t-1}\\
        =& (\alpha_{t|t-1} \alpha_{t-1|s}) x_{s} + \sqrt{ \alpha^2_{t|t-1} \sigma^2_{t-1|s} + \sigma^2_{t|t-1}} \hat{\epsilon}_{s}\\
        =& \alpha_{t|s} x_s + \sigma_{t|s} \epsilon_s & \textcolor{gray}{ x_{t} \sim  q(x_{t} | x_{s})}\\
    \end{aligned}
 \end{equation} 

We conclude that $\alpha_{t|s} = \alpha_{t|t-1} \alpha_{t-1|s}$ and $\sigma_{t|s} = \sqrt{ \alpha^2_{t|t-1} \sigma^2_{t-1|s} + \sigma^2_{t|t-1}}$.

For $\alpha_{t|s} = \alpha_{t|t-1} \alpha_{t-1|s}$,

\begin{equation}
  \begin{aligned}
    \alpha_{t|s} &= \alpha_{t|t-1} \alpha_{t-1|s} \\
                  &= \alpha_{t|t-1} \alpha_{t-1|t-2} \alpha_{t-2|s} \\
                  &\cdots\\
                  &= \prod_{i=s+1}^t \alpha_{i|i-1}\\
                  &= \prod_{i=s+1}^t \frac{\alpha_{i}}{\alpha_{i-1}}\\
  \end{aligned}
\end{equation}

For $\sigma_{t|s} = \sqrt{ \alpha^2_{t|t-1} \sigma^2_{t-1|s} + \sigma^2_{t|t-1}}$,  let $s=0$, then $\sigma^2_{t|t-1} = \sigma^2_{t} - \alpha^2_{t|t-1} \sigma^2_{t-1} $. Therefore,

% \begin{equation}
%   \begin{aligned}
%     \sigma^2_{t|s} &= \alpha^2_{t|t-1} \sigma^2_{t-1|s} + \sigma^2_{t|t-1}
%   \end{aligned}
% \end{equation}

\begin{equation}
  \begin{aligned}
    \sigma^2_{t|s} &= \alpha^2_{t|t-1} \sigma^2_{t-1|s} + \sigma^2_{t|t-1}\\
                    &= \alpha^2_{t|t-1} (\alpha^2_{t-1|t-2} \sigma^2_{t-2|s} + \sigma^2_{t-1|t-2}) + \sigma^2_{t|t-1}\\
                    & = \alpha^2_{t|t-1}\alpha^2_{t-1|t-2} \sigma^2_{t-2|s}   + \alpha^2_{t|t-1}\sigma^2_{t-1|t-2} + \sigma^2_{t|t-1}\\
                    & = \alpha^2_{t|t-2}\sigma^2_{t-2|s}   + \alpha^2_{t|t-1}\sigma^2_{t-1|t-2} + \alpha^2_{t|t} \sigma^2_{t|t-1}\\
                    & = \sum_{k=s+1}^{t}{\alpha^2_{t|k} \sigma^2_{k|k-1}}\\
                    & = \sum_{k=s+1}^{t}{\alpha^2_{t|k} (\sigma_k^2 - \alpha^2_{k|k-1} \sigma^2_{k-1})}  & \textcolor{gray}{\text{Apply } \sigma^2_{t|t-1} = \sigma^2_{t} - \alpha^2_{t|t-1} \sigma^2_{t-1} }\\
                    & = \sum_{k=s+1}^{t}{\alpha^2_{t|k}\sigma_k^2} - \sum_{k=s+1}^{t}{\alpha^2_{t|k-1}\sigma_{k-1}^2}\\
                    & = \alpha^2_{t|t}\sigma_t^2 - \alpha^2_{t|s}\sigma_s^2\\
                    & = \sigma_t^2 - \alpha^2_{t|s}\sigma_s^2
  \end{aligned}
\end{equation}

\end{proof}

 \paragraph{Reverse process} As to the backward process $x_T \rightarrow x_{T-1} \rightarrow \cdots \rightarrow x_0$, the neural network aims to maximize $q(x_{s}| x_{t}, x_0) = \mathcal{N}(z_{s}; \hat{\mu}_s, \hat{\sigma}_s^2 I)$, where 

 \begin{equation}
  \begin{cases}
    \hat{\sigma}_s = \frac{\sigma_{t|s}\sigma_{s}}{\sigma_{t}}\\
    \hat{\mu}_s = \frac{1}{\alpha_{t|s}} x_t - \frac{\alpha_{t|s}}{\sigma_t} \epsilon_t
  \end{cases}
\end{equation}

\begin{proof}

 Recall that $\mathcal{N}(z; \mu, \sigma I) \propto \exp{ \left( -\frac{||z - \mu||^2}{2 \sigma^2} \right)}$, the backward translation could be derived by Bayes' Theorem:
 
 \begin{equation}
   \begin{aligned}
      q(x_{s}| x_{t}, x_0) &= \frac{q(x_{t}| x_{s}) q(x_{s}|x_0)}{q(x_{t}|x_0)}\\
      &\propto \exp{ \left[ -\frac{1}{2} \left( \frac{||x_t - \alpha_{t|s} x_s||^2}{\sigma_{t|s}^2} +  \frac{||x_s - \alpha_{s} x_0||^2}{\sigma_s^2} - \frac{||x_t - \alpha_{t} x_0 ||^2}{\sigma_t^2} \right) \right]}
   \end{aligned}
 \end{equation}
 
 from which we can derive $q(z_{s}| z_{t}, x_0) = \mathcal{N}(z_{s}; \hat{\mu}_s, \hat{\sigma}_s^2 I)$, and
 
 \begin{equation}
   \begin{cases}
     \frac{1}{\hat{\sigma}_s^2} = \frac{\alpha_{t|s}^2}{\sigma_{t|s}^2} + \frac{1}{\sigma_s^2}\\
     -2\frac{\hat{\mu}_s}{\hat{\sigma}_s^2} = -2\frac{x_t\alpha_{t|s}}{\sigma_{t|s}^2} -2\frac{\alpha_s x_0}{\sigma_s^2}
   \end{cases}
 \end{equation}
 
 finally
 
 \begin{equation}
   \begin{cases}
     \hat{\sigma}_s = \frac{\sigma_{t|s}\sigma_{s}}{\sigma_{t}}\\
     \hat{\mu}_s = \frac{1}{\alpha_{t|s}} x_t - \frac{\alpha_{s|t} \sigma^2_{t|s}}{\sigma_t} \epsilon_t 
   \end{cases}
 \end{equation}

% 推导 \hat{mu}_s

Note that
 \begin{equation}
  \begin{aligned}
    -2\frac{\hat{\mu}_s}{\hat{\sigma}_s^2} =& -2\frac{(x_t)\alpha_{t|s}}{\sigma_{t|s}^2} -2\frac{(\alpha_s x_0)}{\sigma_s^2}\\
    \hat{\mu}_s =& \frac{(x_t)\alpha_{t|s}}{\sigma_{t|s}^2} (\frac{\sigma_{t|s}\sigma_{s}}{\sigma_{t}})^2 +\frac{(\alpha_s x_0)}{\sigma_s^2} (\frac{\sigma_{t|s}\sigma_{s}}{\sigma_{t}})^2\\
    \hat{\mu}_s =& \frac{(x_t)\alpha_{t|s} \sigma_{s}^2 }{\sigma_{t}^2}  +\frac{(\alpha_s x_0)\sigma_{t|s}^2}{\sigma_{t}^2}\\
    \hat{\mu}_s =& \frac{(x_t)\alpha_{t|s} \sigma_{s}^2 }{\sigma_{t}^2} +\alpha_s \frac{x_t-\sigma_t \epsilon_t}{\alpha_t} \frac{\sigma_{t|s}^2}{\sigma_{t}^2} \\
    \hat{\mu}_s =& (\frac{\alpha_{t|s} \sigma_{s}^2}{\sigma_{t}^2} + \frac{\alpha_s \sigma_{t|s}^2}{\alpha_t \sigma_t^2}) x_t - \frac{\alpha_s \sigma_{t|s}^2}{\alpha_t \sigma_t} \epsilon_t \\
    \hat{\mu}_s =& \frac{1}{\alpha_{t|s}} x_t - \frac{\alpha_{s|t} \sigma^2_{t|s}}{\sigma_t} \epsilon_t 
  \end{aligned}
\end{equation}

% % 从x_t到x_s的另一种形式, 将\epsilon_t替换为\epsilon
% \begin{equation}
%   \begin{aligned}
%     x_s &= (\frac{1}{\alpha_{t|s}} x_t -  \frac{\alpha_{s|t} \sigma^2_{t|s}}{\sigma_t}  \epsilon_t)+\frac{\sigma_{t|s}\sigma_{s}}{\sigma_{t}} \epsilon_s\\
%     &=  \frac{\alpha_s}{\alpha_t} x_t - \frac{\alpha_s \sigma^2_{t|s}}{\alpha_t \sigma_t}\epsilon_t + \frac{\sigma_{t|s}\sigma_{s}}{\sigma_{t}} \epsilon_s\\
%     &= \frac{\alpha_s}{\alpha_t} x_t + \sqrt{(\frac{\alpha_s \sigma^2_{t|s}}{\alpha_t \sigma_t})^2 + (\frac{\sigma_{t|s}\sigma_{s}}{\sigma_{t}})^2} \epsilon\\
%     &= \frac{\alpha_s}{\alpha_t} x_t +\sqrt{(\frac{\alpha_s \sigma^2_{t|s}}{\alpha_t \sigma_t})^2 + (\frac{\sigma_{t|s}\sigma_{s}}{\sigma_{t}})^2} \epsilon\\
%     &= \frac{\alpha_s}{\alpha_t} x_t +\sqrt{\frac{\alpha^2_s \sigma^4_{t|s}  + \alpha^2_t \sigma^2_{t|s}\sigma^2_{s}}{\alpha^2_t \sigma^2_t}} \epsilon\\
%     &= \frac{\alpha_s}{\alpha_t} x_t +\sigma_{t|s} \sqrt{\frac{ \sigma^2_{t|s}  + \alpha^2_{t|s} \sigma^2_{s}}{\alpha^2_{t|s} \sigma^2_t}} \epsilon\\
%     &= \frac{\alpha_s}{\alpha_t} x_t +\sigma_{t|s} \sqrt{\frac{ \sigma^2_t}{\alpha^2_{t|s} \sigma^2_t}}\epsilon & \textcolor{gray}{\text{Apply } \sigma^2_{t|s} = \sigma^2_{t} - \alpha^2_{t|s} \sigma^2_{s}} \\
%     & = \frac{\alpha_s}{\alpha_t} x_t +\frac{\sigma_{t|s}}{\alpha_{t|s}} \epsilon
%   \end{aligned}
% \end{equation}

% Denote $\lambda_t = \frac{1}{(1+\sigma_{t|s})}$, and $C_t = \frac{(1+\sigma_{t|s})}{\alpha_{t|s}}$, then

% \begin{equation}
%   x_s = C_t (\lambda_t x_t  + (1-\lambda_t) \epsilon) \quad \textcolor{gray}{\text{//global mixup}}
% \end{equation}
 
\end{proof}
 
\clearpage
\subsection{Ablation}

\paragraph{Objective\&Setting} We conduct ablation experiments to investigate the impact of conditions and constraints on protein backbone inpainting. Specifically, we show how these factors affect the validation losses, including angle loss, length loss, and overlapping loss.

\begin{figure}[h]
  \centering
  \includegraphics[width=6in]{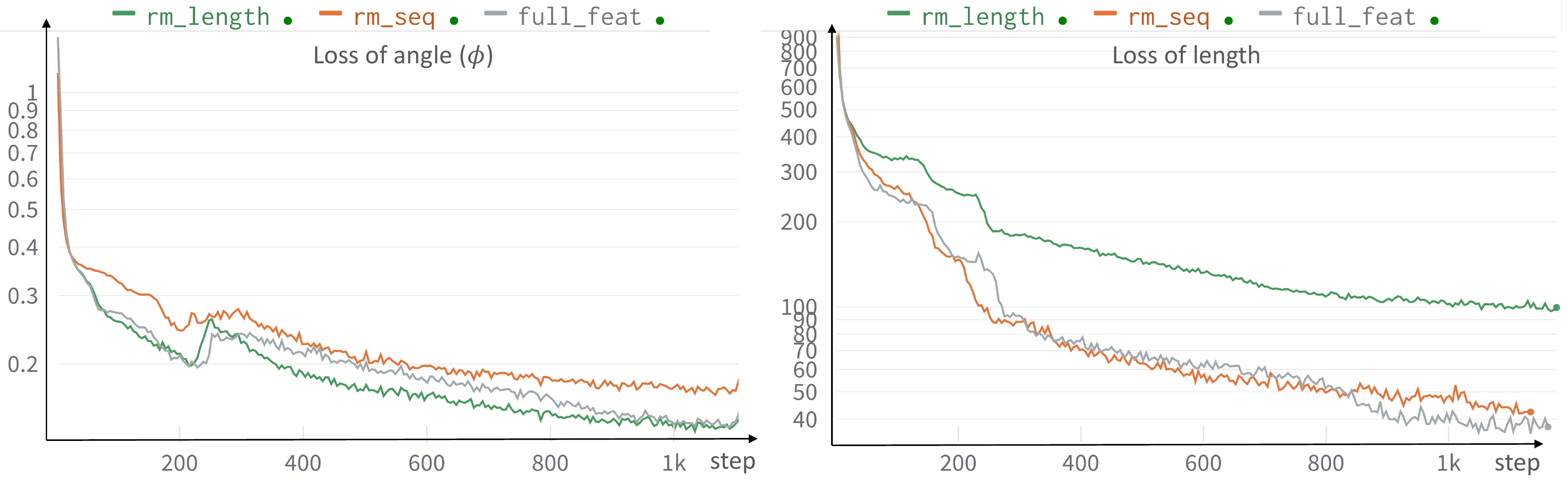}
  \caption{Ablation of conditions. We remove length condition or residue type embedding from the baseline ($\text{full}\_{\text{feat}}$) model, resulting in $\text{rm}\_{\text{length}}$ and  $\text{rm}\_{\text{seq}}$, respectively. We show the loss curves for angle ($\phi$) and length on the validation set, revealing the effect of the conditions. }
  \label{fig:ablation_cond}
\end{figure}

\begin{figure}[h]
  \centering
  \includegraphics[width=6.8in]{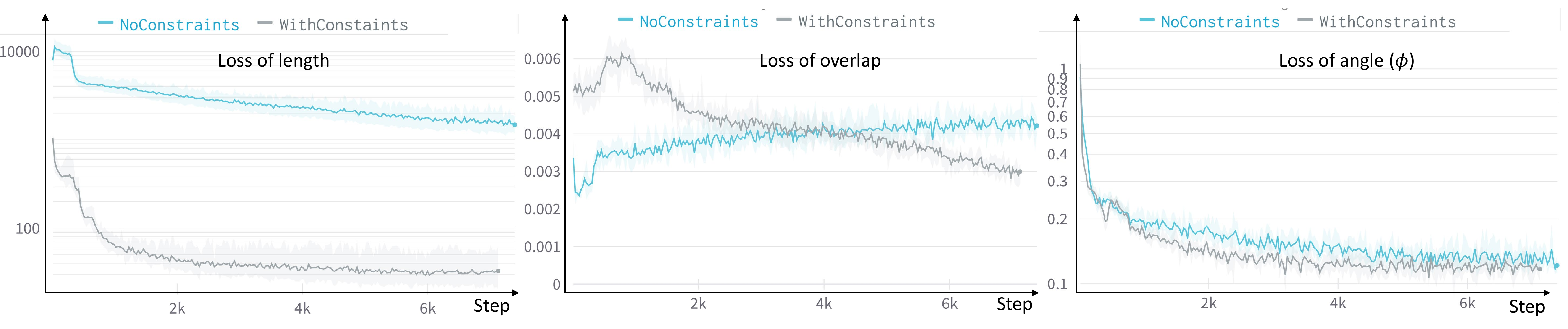}
  \vspace{-4mm}
  \caption{Ablation of constraints. We show the validation loss curves in the constrained and unconstrained cases. The unconstrained case means that $mathcal{L}_{len}$ and $mathcal{L}_{overlap}$ are not imposed on the model during training.}
  \label{fig:ablation_cons}
\end{figure}

\paragraph{Results \& Analysis}  As shown in Fig.\ref{fig:ablation_cond} and Fig.\ref{fig:ablation_cons}, we find that: (1) the length condition and the sequence condition contribute to the reduction of $\mathcal{L}_{len}$ and $\mathcal{L}_{sim}$, respectively. This phenomenon suggests that the model can learn to generate structures with a predetermined length and that residue types can facilitate learning backbone angles. (2) Explicitly imposing geometric constraints on the model is necessary. If the constraints are removed, the $\mathcal{L}_{len}$ loss is difficult to reduce, and the $\mathcal{L}_{overlap}$ even increases, indicating the model could not generate geometrically reasonable structures. Fortunately, all these drawbacks could be eliminated by imposing geometric loss on the introduced direction space.

%  \begin{algorithm} 
%      \caption{Training} 
%      \begin{algorithmic}
%          \FOR {$\text{epoch} = 0, ..., 1000$}
%          \STATE Sampling $t$
%          \STATE 
%          \STATE Compute $\hat{\mu}_{t-1}$ and $\hat{\sigma}_{t-1}$ from Eq.(21)
%          \STATE ${x}_{t-1} =  \hat{\mu}_{t-1} + \hat{\sigma}_{t-1} z$
%      \ENDFOR
%     %  \RETURN $x_0$
%      \end{algorithmic} 
%  \end{algorithm}

%  \begin{algorithm} 
%      \caption{Ancestral Sampling} 
%      \begin{algorithmic}
%      \STATE ${x}_{T} \sim \mathcal{N}({0}, {I})$
%          \FOR {$t = T, ..., 1$}  
%          \IF{$t > 1$} 
%              \STATE $z \sim \mathcal{N}({0}, {I})$ 
%          \ELSE 
%              \STATE $z = 0$ 
%          \ENDIF
         
%          \STATE Compute $\hat{\mu}_{t-1}$ and $\hat{\sigma}_{t-1}$ from Eq.(21)
%          \STATE ${x}_{t-1} =  \hat{\mu}_{t-1} + \hat{\sigma}_{t-1} z$
%      \ENDFOR
%     %  \RETURN $x_0$
%      \end{algorithmic} 
%  \end{algorithm}
 
%  \paragraph{Multiplying Quaternion}
 
%  \begin{equation}
%    \begin{cases}
%      q(R_{ji}) = q(Q_i^T Q_j) = q(Q_i^T) q(Q_j)\\
%      q(R_1^T) = [q_1, -q_2, -q_3, -q_4]
%    \end{cases}
%  \end{equation}

% \begin{figure*}[h]
%   \centering
%   \includegraphics[width=5.3in]{figs/ramachandran.pdf}
%   \caption{Ramachandran plot. }
%   \label{fig:ramachandran}
% \end{figure*}

%%%%%%%%%%%%%%%%%%%%%%%%%%%%%%%%%%%%%%%%%%%%%%%%%%%%%%%%%%%%%%%%%%%%%%%%%%%%%%%
%%%%%%%%%%%%%%%%%%%%%%%%%%%%%%%%%%%%%%%%%%%%%%%%%%%%%%%%%%%%%%%%%%%%%%%%%%%%%%%

\end{document}